\documentstyle[12pt,aaspp4]{article}
\begin{document}

\title{What do the UV Spectra of Narrow-line Seyfert 1 Galaxies tell
us about their BLR?}

\author{Joanna Kuraszkiewicz\altaffilmark{1}, Belinda J. Wilkes} 
\affil{Harvard-Smithsonian Center for Astrophysics, Cambridge, MA
02138}

\author{Bo\.zena Czerny}
\affil{N. Copernicus Astronomical Center, Warsaw,
Poland}

\and

\author{Smita Mathur}
\affil{The Ohio State University, Columbus, OH 43120\altaffilmark{2}}

\altaffiltext{1}{also N. Copernicus Astronomical Center, Warsaw,
Poland}
\altaffiltext{2}{also Harvard-Smithsonian Center for Astrophysics,
Cambridge, MA 02138}

\begin{abstract}

We study the UV spectra of narrow-line Seyfert~1 (NLSy1) galaxies and
compare them with ``normal'' AGN.  Similar to their optical lines, the
NLSy1s show narrower UV lines. They are also characterized by weaker
CIV\,$\lambda$1549, CIII]\,$\lambda$1909, and stronger
AlIII\,$\lambda$1857 emission. These UV line properties add to the
optical and X-ray properties known to be part of the Boroson \& Green
eigenvector 1. We show that the steep soft-X-rays, which characterize
the NLSy1s SEDs, change the equilibrium of the two phase
cloud-intercloud medium resulting in somewhat higher BLR cloud
densities, lower ionization parameter and larger BLR radii. These
modified conditions can explain the unusual emission line properties
we find in NLSy1.

Using a specific model of an accretion disk with corona
presented by Witt, Czerny \& \.Zycki, we also show that the
steep soft and hard-X-ray continua can be explained if the $L/L_{Edd}$
ratios are larger than in ``normal'' Seyfert1/QSO strengthening
earlier suggestions that the $L/L_{Edd}$ is the physical parameter
driving this eigenvector.

\end{abstract}


\section{Introduction}

Narrow-line Seyfert~1 galaxies (NLSy1), first suggested as a distinct
class of AGN by Osterbrock and Pogge (1985), are characterized by
Balmer lines whose FWHM is smaller than typical Seyfert 1 galaxies
i.e.  500 $<$ FWHM $<$ 2000 km s$^{-1}$, slightly broader than the
forbidden lines. On the other hand they are clearly different from
Seyfert 2 galaxies since the ratio of $[$OIII]\,$\lambda$5007 to
H$\beta$ is $<$3, i.e. below the limiting value found by Shuder and
Osterbrock (1981) to discriminate between Seyfert~1 and Seyfert~2
galaxies. In NLSy1 strong Fe II optical multiplets and higher
ionization iron lines (e.g.$[$FeVII]6087\AA\ and $[$FeX] 6375\AA) are
often present. These are usually seen in Seyfert~1 and not in
Seyfert~2 galaxies.

Many NLSy1s have an unusually strong big blue bump (BBB) which, when
compared to typical Seyfert~1 and QSO BBBs, is shifted towards higher
energies, sometimes even out of the optical/UV range (at least one
object actually peaks in the soft X-ray band: RE~J1034+396,
Puchnarewicz et al. 1995).  Its high frequency tail is clearly seen in
soft X-rays and these objects have generally steeper soft X-ray
continua than is ``typical'' for Seyfert~1 galaxies (Boller, Brandt \&
Fink 1996), meaning that they have a stronger soft-X-ray excess over
the hard X-ray power law. The intrinsic hard-X-ray continua of NLSy1s
are also generally steeper (Brandt, Mathur, Elvis 1997) than in
typical Seyfert~1s. The NLSy1s are usually only weakly absorbed in the
soft X-rays (Boller, Brandt \& Fink 1996) and in many cases both the
UV flux and soft X-ray flux are strongly variable.

NLSy1 objects are generally radio-quiet and their radio powers are
typical of those found in other Seyfert galaxies (Ulvestad, Antonucci
\& Goodrich 1995).

There is no widely adopted view on the basic reason why the continua of
NLSy1 galaxies are different from classical Sy1. The two most probable
explanations of the stronger big blue bumps in these objects are pole-on
orientation (Puchnarewicz, Mason, C\'{o}rdova 1994, Wilkes 1998), and
higher accretion rate relative to the mass of the central object (e.g.
Boller, Brandt, Fink 1996; Wandel 1997, Czerny, Witt \& \. Zycki 1997,
Pounds, Done \& Osborne 1995).  Steeper hard X-ray spectra and strong
permitted FeII lines are possibly a secondary effect of the atypical
shape of the soft-X-ray continuum (Pounds, Done \& Osborne 1995, 
Brandt, Mathur \& Elvis 1997; Wilkes, Elvis and McHardy 1987, Shastri
et al 1993).


Wilkes et al. (1999), studying a sample of low
redshift quasars and Sy1s, and their relations between optical/UV
emission lines and the continuum, found that the four NLSy1 in their
sample show smaller equivalent widths of CIII] and CIV lines than
typical AGN (EW(CIV) $<$ 40\AA\ for NLSy1, while 30\AA\ $<$ EW(CIV)
$<$ 200\AA\ for other AGN). In this paper we investigate in detail the
UV line properties of a sample of NLSy1 objects to determine whether
the weakness of the carbon lines is typical of these objects,
constituting an additional property which distinguishes them from
``normal'' Sy1 galaxies. We investigate the physical conditions of the
BLR which may explain these systematic differences. We also discuss
the possibility that these objects have luminosities close to their
Eddington luminosity.

\section{UV line measurements}

\subsection{The sample}

From the currently known set of NLSy1 (Boller, Brandt \& Fink 1996,
Greiner et al. 1996, Puchnarewicz et al. 1992, Puchnarewicz et
al. 1994, Brandt, Fabian \& Pounds 1996, Grupe et al. 1996 , Moran et
al. 1996, Brandt - private communication, Wilkes et al. 1999) we have
defined a subset of 11 objects (Table~1) for which UV spectra are
available either from the HST (5 objects) or IUE archives.  The IUE
spectra were taken from Lanzetta et al. (1993), and the reduced HST
spectra from Dobrzycki (private communication, see also Bechtold et
al. 2000).

\subsection{Line parameters}

We have measured the EW (Table~2), line ratios (Table~3) and line
widths (FWHM; Table~4) of all prominent UV lines: Ly$\alpha$\,$\lambda$1216,
CIV\,$\lambda$1549, CIII]\,$\lambda$1909, SiIII]\,$\lambda$1892,
AlIII\,$\lambda$1857, SiIV+OIV]\,$\lambda$1400 blend, and
MgII\,$\lambda$2798. 

The EW and FWHM of IUE spectra were measured using the {\it splot}
task in IRAF: the EW by fitting a linear continuum to the data and
integrating across the observed emission line (keystroke `e`),
the FWHM by measuring the width at half the flux in the line peak
above the continuum. The same procedure was applied when the line
parameters in the HST spectra were measured, although a different
program {\it findsl} (provided by Aldcroft, Bechtold \& Elvis 1994),
specially written to handle the HST data, was used. The line
parameters presented in Tables~2,3,4 have been corrected for
absorption: for weak absorption by using a linear fit across the
absorption line, for strong absorption (as in PG~1351+560 and
PG~1411+442) by assuming a symmetric emission line profile and
reflecting the unabsorbed wing about the peak.


As has been noted by Vestergaard \& Wilkes (2000), CIII] is blended
with FeIII UV34 $\lambda$1914 line, which should be taken into account
especially when the CIII] line is weak. In three spectra (IZw1,
PG~1211+143, Mrk~478), where the FeIII UV34 was clearly visible, we
subtracted this line (modeled as a Gaussian centered at
$\lambda$=1914{\AA} rest frame) from the CIII] blend.

\subsection{Comparison of NLSy1 with ``normal'' AGN}

In this section we compare the UV line properties of NLSy1 with
Seyfert~1 galaxies and quasars.  Fig.~1a shows the EW of Ly$\alpha$,
CIV and MgII of our NLSy1 sample (shaded areas) compared to the sample
of Seyfert~1 galaxies (dotted line) from Wu at al. (1983) (their
sample includes three NLSy1: IZw1, Mrk 478, IIZw136 which we excluded
here) and low redshift quasars from Wilkes et al. (1999), Corbin \&
Boroson (1996), and radio-loud quasars from Baldwin, Wampler \&
Gaskell (1989), combined and denoted by a dashed line.  The EW of CIV
and MgII lines are significantly smaller in NLSy1 than in the broad
line Seyfert 1 galaxies and quasars. The K-S test yielded a 0.001
chance that the EW of CIV and MgII in NLSy1s and Sy1s are drawn from
the same population. For Ly$\alpha$ the chance was $<$ 0.02. When
compared to QSO the significance remained strong for CIV ($p <
0.01$) and MgII ($p < 0.025$), while for Ly$\alpha$ the
distributions are similar ($p > 0.5$). The smaller EW of the carbon
and MgII lines cannot be due to a simple continuum increase, as this
would effect the EW of all lines equally, while EW(Ly$\alpha$) is not
significantly smaller.

\begin{figure} [t!]
\vspace{9.0truecm}
{\includegraphics{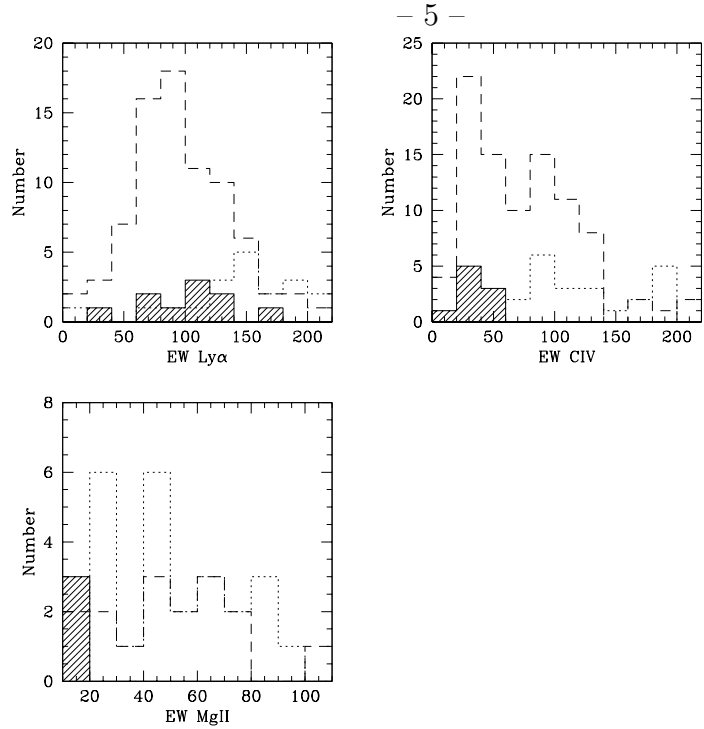}}
\caption{a) Comparison of Ly$\alpha$, CIV, and MgII
equivalent widths in our NLSy1 sample (shaded areas) with the
Seyfert~1 sample from Wu et al. (1993) (dotted line) and QSOs (dashed
line) from Wilkes et al. (1998), Corbin \& Boroson (1996), and
Baldwin, Wampler \& Gaskell (1989)}
\end{figure}

As can be seen from Table~3 the CIV/Ly$\alpha$ (mean 0.25$\pm$0.09),
CIII]/Ly$\alpha$ (mean 0.05$\pm$0.05) and the MgII/Ly$\alpha$
(mean 0.05$\pm$0.03) ratios are smaller compared to those typically
observed in Seyfert 1 galaxies (Wu et al. 1983 give observed ranges: 
0.35-2.01, 0.03-0.39, 0.07-0.63 respectively) and quasars (observed
range: 0.3-1.04, 0.15-0.3, 0.15-0.35). To show this more clearly
Figure~2 shows the CIII]/Ly$\alpha$ vs CIV/Ly$\alpha$ line ratios
for NLSy1 in our sample (denoted as filled squares) with the Seyfert~1
sample (denoted as circles) and quasars from Laor et al. (1995),
Christiani \& Vio (1990), Wilkes et al. (1999) and narrow line quasars
from Baldwin et al. (1988).

As the lines in NLSy1 are narrow, we can clearly resolve the
components of the CIII]+SiIII]+AlIII blend (especially in the HST
data, in the IUE data the S/N is often too low). From Table~3 it is
also clear that the SiIII] line in most of the NLSy1 is very strong
compared to the CIII] line. Also the SiIV+OIV] blend is strong
compared to CIV (mean SiIV+OIV]/CIV ratio in NLSy1 is 0.49$\pm$0.26,
larger than the mean ratio of 0.3 in quasars from Francis et
al. 1991). However, the SiIV+OIV]/Ly$\alpha$ ratio is in the range of
normal AGN, indicating that the large SiIV+OIV]/CIV ratio is due to
weaker CIV emission. The AlIII doublet in NLSy1 is rather
strong (equivalent width $\sim$ few \AA\ - see Table~2).

Although broader than H$\beta$ the UV lines in NLSy1 are narrow
compared to other AGN. In Table~4 and Fig.~1b we compare our objects
with samples of low redshift quasars from Corbin \& Boroson (1996) and
Wilkes et al. (1999), and with a radio-loud sample from Baldwin,
Wampler \& Gaskell (1989).

\begin{figure} [t!]
\vspace{9.0truecm}
{\includegraphics{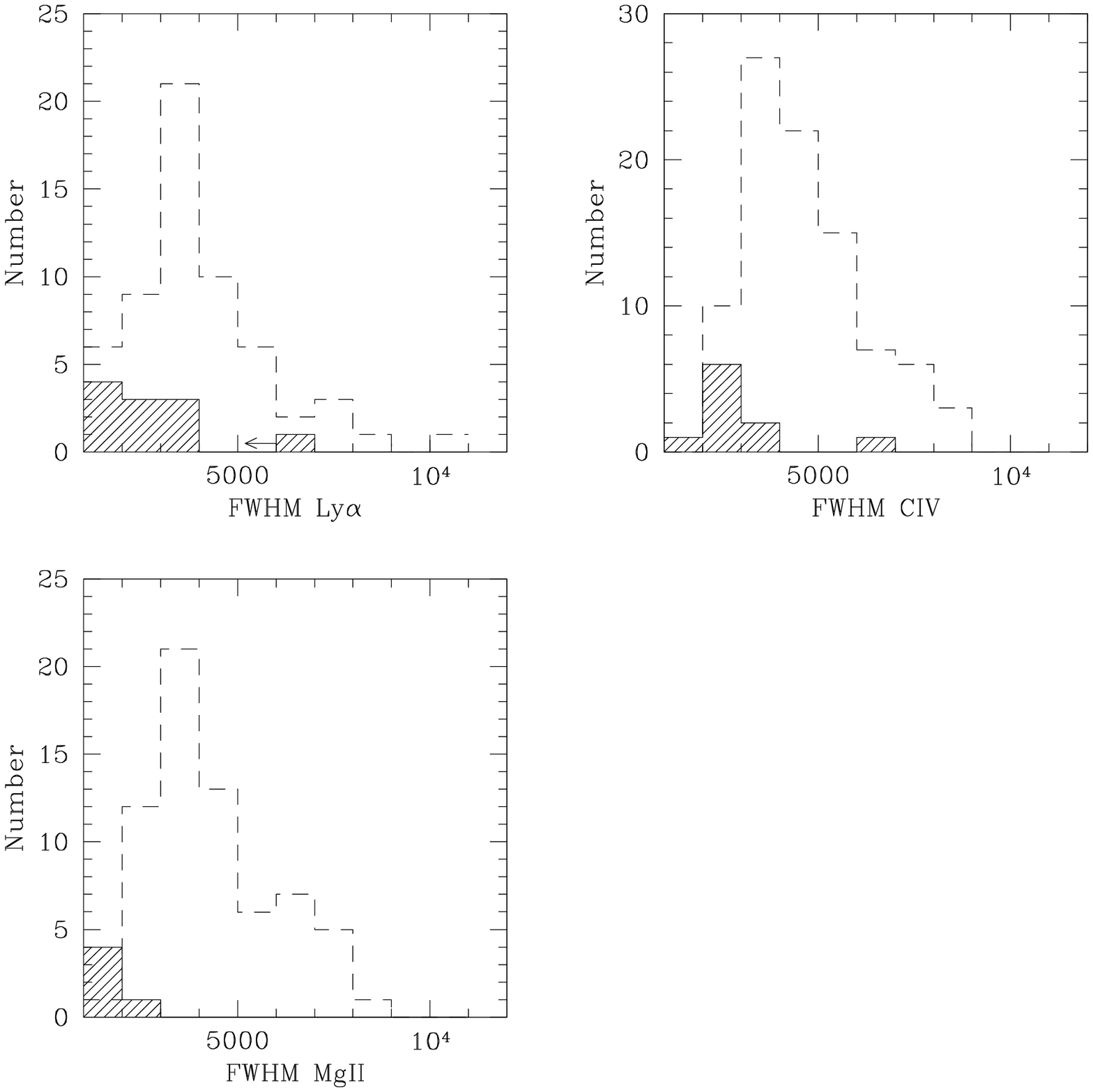}}
\addtocounter{figure}{-1}
\caption{b). Comparison of FWHM of
Ly$\alpha$, CIV, and MgII full widths at half maximum, with the same
coding as in Fig 1a.}
\end{figure}


\begin{figure} [t!]
\vspace{9.0truecm}
{\includegraphics{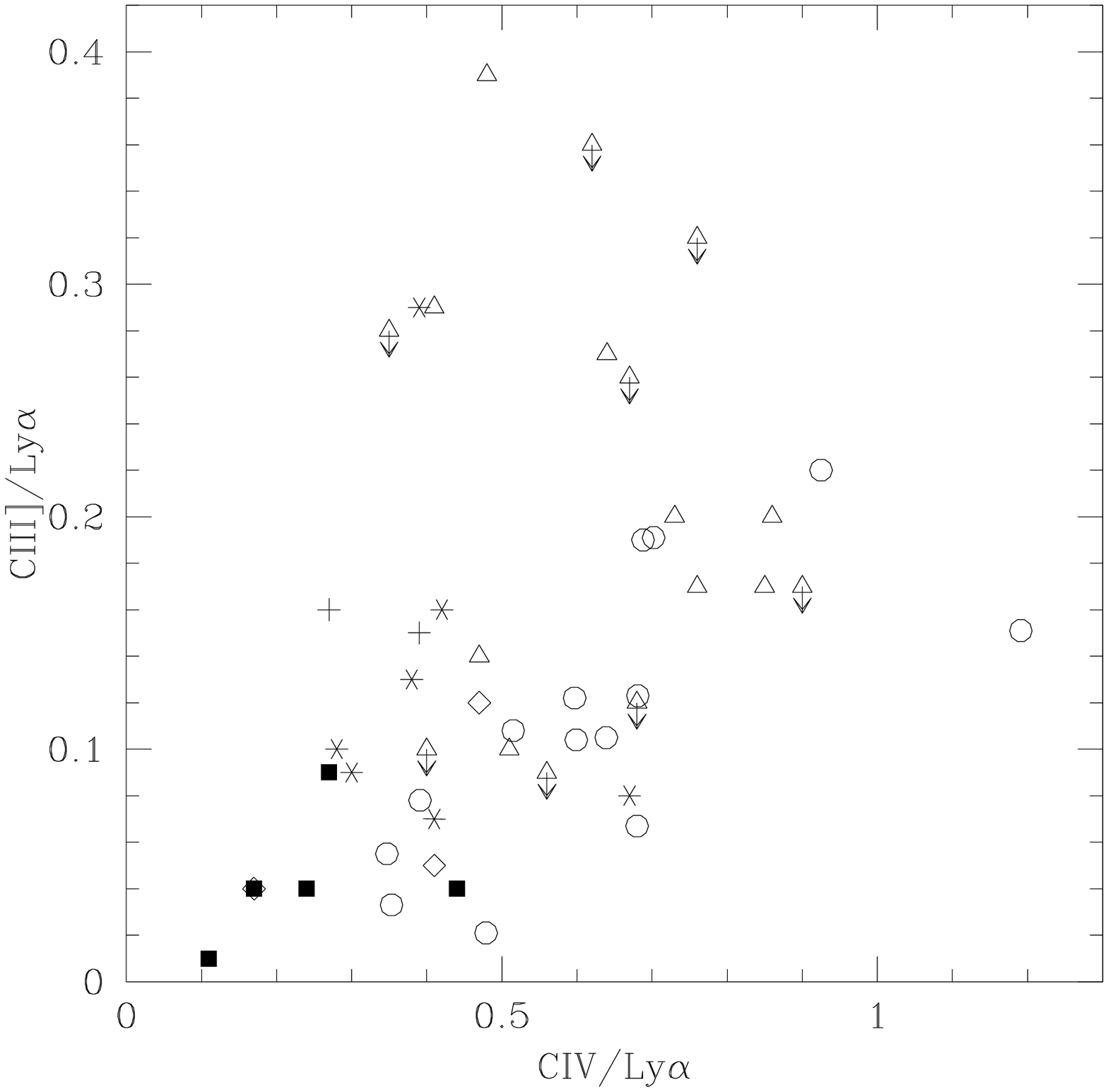}}
\caption{Comparison of CIII]/Ly$\alpha$ and CIV/Ly$\alpha$
ratios for NLSy1 objects (filled squares) and Seyfert~1 galaxies from
Wu et al. (1983, circles) and QSOs from Laor et al. (1995, stars),
Christiani \& Vio (1990, crosses), Wilkes et al. (1998, triangles) and
narrow line quasars from Baldwin et al. (1988, diamonds).  (For CIII]
we used the sum of CIII]+SiIII]+AlIII to allow comparison with other
samples where the broader lines prevented the authors from separating 
these components.)}
\end{figure}

\section{Discussion}

We will now investigate what the line strengths and line ratios tell
us about the physical properties in the BLR clouds of NLSy1 (Section
3.1).  
Then we will discuss the continuum properties of NLSy1 (Section 3.2)
and investigate what they indicate about their central engine. In
conclusion we show how the deduced differences between the central
engines of NLSy1 and ``normal'' AGN can explain their different,
observed emission line spectra.

\subsection{Physical properties of the BLR clouds in NLSy1}

More than ten years ago Gaskell (1985) noticed that Seyfert 1 galaxies
with narrow H$\beta$ lines of FWHM $<$ 1600 km s$^{-1}$ show lower
H$\beta$ equivalent widths than typical Seyfert 1. He interpreted
this finding as a result of collisional destruction of H$\beta$ in the
higher density BLR clouds in these objects. Although we do not study
the optical spectra of NLSy1 in this paper, we will now investigate 
whether the UV spectra lead to a similar conclusion.

Rees et al. (1989) calculated the line intensities for different BLR
cloud densities at constant column density ($10^{23}$ cm$^{-2}$) and
ionization parameter ($U=10^{-2}$). They found that optically thick
lines such as hydrogen and carbon lines have a fairly constant
intensity up to a certain density, above which these lines become
thermalized and their intensity drops considerably. For hydrogen
lines, CIV and MgII this critical density is $\sim 10^{10}$cm$^{-3}$.
For the semi-forbidden lines CIII] and SiIII] it is around $5\times
10^{9}$ cm$^{-3}$ and $10^{11}$ cm$^{-3}$ respectively. The Ly$\alpha$ and
CIV lines are usually strong coolants at densities smaller than these
critical values, but as the density increases and these lines become
thermalized, other high-excitation lines such as CIII\,$\lambda$977 and
AlIII\,$\lambda$1857 take over the cooling.

In the previous section we showed that the UV spectra of NLSy1s, when
compared to ``normal'' Seyfert~1 and QSO galaxies, show weaker carbon
and MgII lines. Although the wavelength of our spectra does not cover
the range of CIII\,$\lambda977$, the AlIII\,$\lambda$1857 doublet is
clearly seen (where the S/N is high enough) and is especially strong
in IZw1 (see Table~2).  All these line properties suggest that in
NLSy1 objects the BLR clouds have higher densities than the BLR clouds
in ``normal'' AGN. We will estimate how much higher by studying the
line ratios in the following section.


\begin{center}
{\it {a) The line ratios}}
\end{center}

The CIII] and CIV to Ly$\alpha$ ratios are often used as a density
indicator.  This is because the carbon lines are collisionally excited
(hence sensitive to density), while Ly$\alpha$ is not (note however
that Mathur et al. 1994 showed that for large ionizing parameters, $U
> 0.1$, where $U = \frac
{\int_{1Ryd}^{\infty}\frac{L_{\nu}}{h\nu}}{4\pi r^{2}cn_{H}}$, CIII]
ceases to be a density indicator). These line ratios are also a
sensitive function of the ionization parameter U, therefore we
investigate the relation of these line ratios to density and U.

We have calculated line ratios using the photoionization code CLOUDY
(version 80.07, for reference see Ferland 1991).  First as an input
ionizing continuum we took a standard AGN continuum (table agn;
Mathews \& Ferland 1987). With this continuum, the observed lines were
only reproduced with higher densities in the CIII] than the CIV
emitting clouds. This requires a steep increase in cloud density with
radius, which is contrary to expectations and seems unrealistic.  Then
we used the spectral energy distribution (SED) of the NLSy1
PG~1211+143 as the ionizing continuum. Although no detailed study of
the SEDs of NLSy1s has been made and is beyond the scope of this
paper, PG~1211+143 is typical of those studied to date with
$\alpha_{x}=2.13\pm0.22$, where $F_{\nu}=\nu^{-\alpha}$ (Wang,
Brinkmann \& Bergeron 1996; where typical NLSy1 slopes are in the
range 1.5 to 3.5 - see Boller, Brandt \& Fink 1996) and
$\alpha_{io}=0.90$ (i.e. slope measured between 1 $\mu$m and 2500\AA,
where typical NLSy1 values are 0.4 to 2.8 - see Lawrence et
al. 1997). The IR to hard X-ray SED of this object was taken from
Elvis et al. (1994) and is reproduced here in Fig.~3. The SED was
linearly interpolated between the observational points in the
optical/UV region. The EUV continuum was determined by a linear
interpolation between the lowest energy point in the X-ray range and
the highest in the UV, providing a conservative (i.e. low) estimate of
the number of EUV photons. We investigated a range of cloud densities
($n(H) = 10^8 - 10^{13}$ cm$^{-3}$, well within the range of the
applicability of the photoionization code 
CLOUDY\footnote{``The hydrogen and atoms and ions of helium are
treated in the code as 10-level atoms. The treatment of the heavy
elements is not as complete as hydrogen and helium, but a 3-body
recombination is included as a general recombination process. [...]
The physical high-density limit is set by the approximate treatment
of the three-body recombination-collisional ionization ($ \le$
$10^{13}$ cm$^{-3}$) for the heavy elements and the approximate
treatment of line transfer''. - see Ferland (1991).}) and ionization
parameters ($U = 10^{-3} - 10^{-1}$, where $U=10^{-2}$ is the value
for the ``standard'' BLR - Davidson \& Netzer 1979).  The metal
abundances were assumed to be solar and the cloud column density
10$^{23}$ cm$^{-2}$. The calculated line ratios are plotted in Fig.~4,
where the observed line ratios for our NLSy1 are denoted by horizontal
lines.  The CIV/Ly$\alpha$ ratio (Fig.~4a) depends very strongly (more
than any other line ratio) on the value of the ionization parameter as
well as on the density of the BLR clouds. Small values of the
ionization parameter ($U = 10^{-3}$) are clearly favored by our data,
for which densities of the order of 10$^{11}$ cm$^{-3}$ to 10$^{12}$
cm$^{-3}$ are needed to produce the low observed CIV/Ly$\alpha$
ratios. For the same, small ionization parameter the observed
CIII]/Ly$\alpha$ ratios indicate densities between $ > 10^{9}$ cm$^{-3}$
and 10$^{11}$ cm$^{-3}$ (see Fig.~4b), which are smaller than the cloud
densities inferred from the CIV/Ly$\alpha$ ratios. This strongly
suggests that the CIII] and CIV lines are formed in different clouds,
which implies a stratified BLR.  This is as expected from the results
of reverberation mapping (e.g. NGC~5548 Korista et al. 1995, NGC~7469
Wanders et al. 1997, Fairall 9 Rodrigues-Pascual et al. 1997) which
show that the CIV and Ly$\alpha$ line fluxes vary with a smaller time
delay, relative to the UV continuum, than the CIII] lines, indicating
that CIV, Ly$\alpha$ emitting clouds lie nearer to the central engine
than the CIII] emitting clouds. The density of the CIV and Ly$\alpha$
emitting line region is typically estimated (e.g. Peterson et al. 1985)
to be $\sim$ 10$^{11}$ cm$^{-3}$, while the CIII] region is $\sim$
10$^{9.5}$ cm$^{-3}$. Thus the density of the CIV, Ly$\alpha$ 
emitting clouds in our NLSy1s is comparable or somewhat ($<$ 10 times)
larger than in normal AGN.

\begin{figure} [t!]
\vspace{9.0truecm}
{\includegraphics{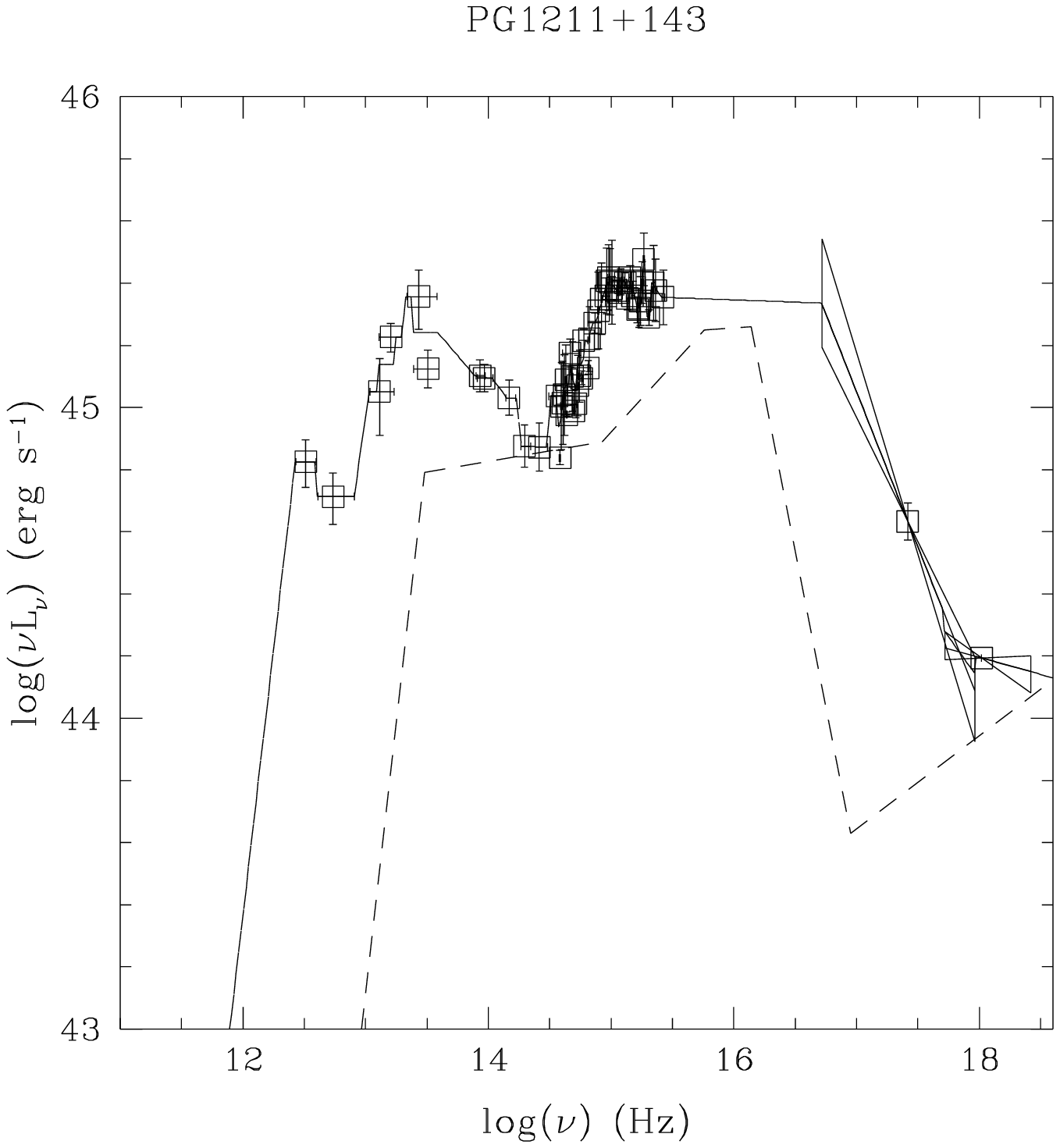}}
\caption{The infrared to hard X-ray spectral energy distribution
of PG~1211+143 (from Elvis et al. 1994), which was used as an input 
ionizing continuum in the photoionization calculations. The dotted line
indicates the standard AGN continuum of Mathews \& Ferland (1987)
used in CLOUDY normalized at 1$\mu$m. The NLSy1s have a more
pronounced BBB and a strong soft-X-ray excess.}
\end{figure}

\begin{figure} [t!]
\vspace{9.0truecm}
{\includegraphics{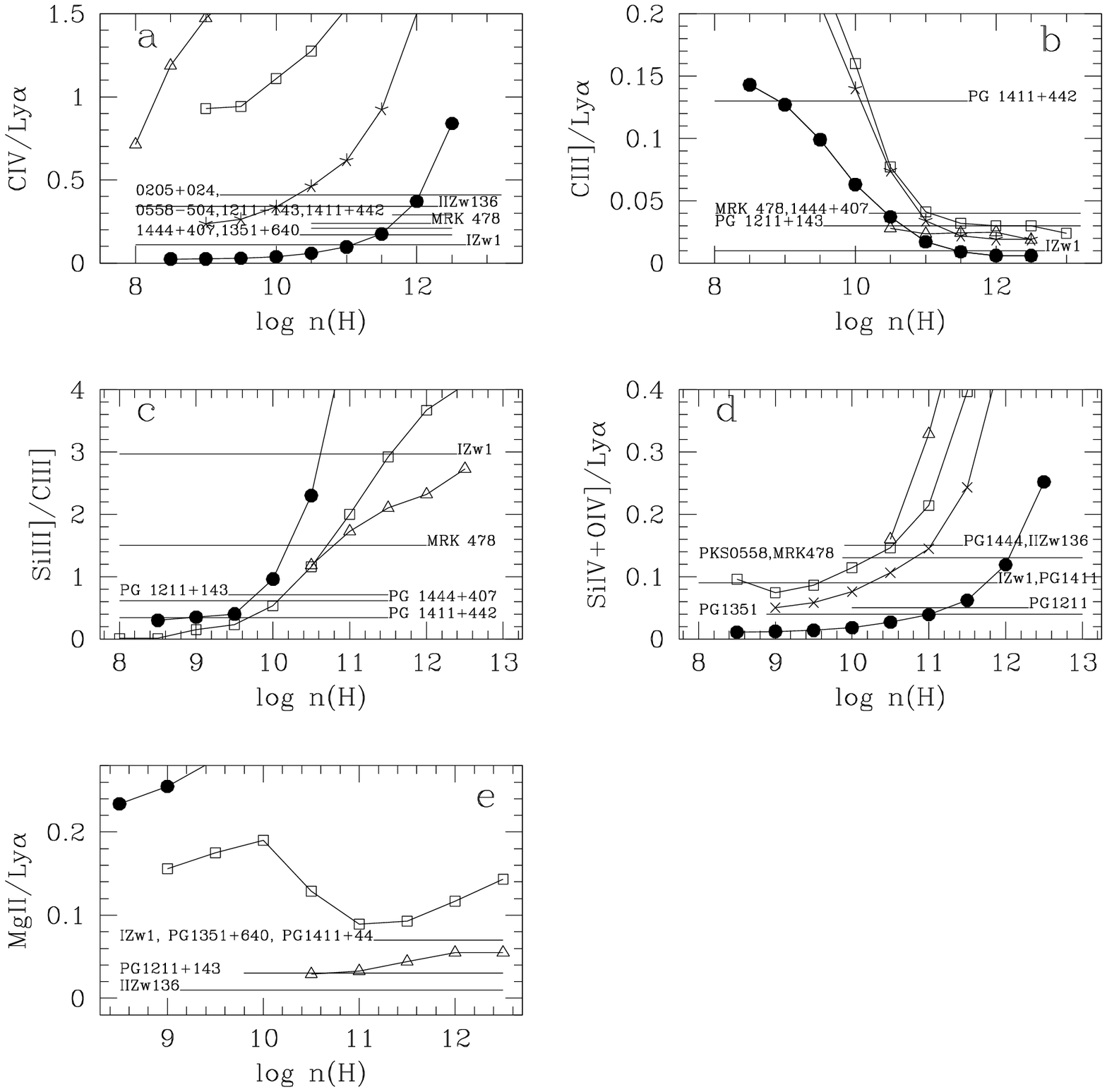}}
\vspace{1.5cm}
\caption{Calculated line ratios of a) CIV/Ly$\alpha$, b)
CIII]/Ly$\alpha$, c) SiIII]/CIII], d) SiIV+OIV]/Ly$\alpha$, e)
MgII/Ly$\alpha$,  for several different
ionization parameters. Triangles denote $\log U=-1$, squares 
$\log U=-2$, filled
circles $\log U=-3$ and stars $\log U=-2.5$. Horizontal lines show the
observed line ratios.}
\end{figure}

To further constrain the density of the CIII] emitting clouds, we need
to investigate its ratio to a line which is formed in the same clouds
for example SiIII]. For $U = 10^{-3}$ the density of the emitting gas
inferred from the SiIII]/CIII] line ratio is between 10$^{9.5}$
cm$^{-3}$ and 10$^{10.5}$ cm$^{-3}$ (see Fig.~4c, we omit here
PG~1411+442, which is a BAL QSO and has $n(H) \sim 10^9$
cm$^{-3}$). The ratio of SiIII]/CIII] is larger than the typical value
of $\sim 0.3 \pm 0.1$ seen in quasars (Laor et al. 1995) for all
objects (except for BAL QSO PG~1411+442; see Table~3 and Fig.~4c) 
This high ratio is probably the result of the suppression of CIII]
while SiIII] remains strong, due to the smaller critical density for
CIII] ($ \ge 5 \times 10^{9}$ cf.  $10^{11}$ cm$^{-3}$ for SiIII], see
Section 3.1a).

In Fig.~4d we present the observed and calculated SiIV+OIV]/Ly$\alpha$
ratios. For $U = 10^{-3}$ the clouds have density of the order of
10$^{11}$ cm$^{-3}$ to 10$^{12}$ cm$^{-3}$, similar to the range for
CIV and Ly$\alpha$ emitting clouds.


The MgII/Ly$\alpha$ ratio, on the other hand, is very small, and
cannot be reproduced by clouds with small ionization parameter even
when the column density, $N_{H}$, is varied over the range $10^{22}$ -
$10^{24}$ cm$^{-2}$. Only larger than standard ionization parameters,
in the range $10^{-1}$ to $10^{-2}$, can produce the observed ratios
(Fig.~4e). This may indicate that the MgII lines do not form in the
same region as the other BLR lines. We will return to this problem in
the next section.



To summarize the results of this subsection we conclude that the
unusual UV line ratios in the NLSy1 objects can be explained if the
BLR clouds have 10 times lower ionizing parameters ($\log U \sim -3$)
and a few times ($<$10) higher densities (n(H)$\sim$ 10$^{11-12}$
cm$^{-3}$ for Ly$\alpha$, CIV, SiIV emitting clouds and
10$^{9.5-10.5}$ cm$^{-3}$ for CIII], SiIII] emitting clouds) than
normal AGN. The BLR is clearly stratified with CIV, Ly$\alpha$, SiIV
producing clouds lying closer to the central engine, while CIII] and
SiIII] emitting clouds lying predominantly further out. The MgII
emission cannot be produced by the same cloud population, suggesting
that these lines form in a different region.

Although clouds with a wide range of properties are likely to exist in
the broad-line region, it was shown by Baldwin et al. (1995) and
Korista et al. (1996), that each emission line is most efficiently
produced in gas with the optimum parameters for that line. These are
the so called locally optimally emitting clouds or LOCs. Thus our
modeling derives the parameters of the LOCs for each line so that the
line fluxes and ratios provide a good approximation to a detailed
multi-zone model of the BLR (see Baldwin et al 1995, Korista et
al. 1996 and our Table~3), which is beyond the scope of this paper.

\begin{center}
{\it {b) The weak MgII problem}}
\end{center}


The MgII line is surrounded on both sides by FeII emission.  As the
FeII emission in NLSy1s is usually very strong (Boller, Brandt \& Fink
1996), it is possible that the wings of MgII disappear in the stronger
iron bumps. This effect could lead to an underestimation of the MgII
emission of up to a factor of two in the strongest FeII sources (such
as IZw1 - see Vestergaard and Wilkes 2000). However, even if our
measurements were underestimating the MgII emission by such a large
factor, the real MgII/Ly$\alpha$ ratio would still be much smaller
than that observed in quasars or in the lower end of the range for
Seyfert~1s (see Table~3). The ionization parameter U inferred from the
observed line ratio would be $\sim 10^{-1} - 10^{-2}$ (see Fig.~4e),
still larger than that inferred from the other emission lines.

Thus we conclude that the MgII emitting clouds have a different value
of the ionization parameter, and are formed in a physically different
region of the BLR (consistent with the different time lags shown by
Ly$\alpha$, CIV and MgII lines in NGC~5548). Photoionization
models predict that MgII is either formed in a partially ionized zone
(PIZ) of the BLR clouds or in a low ionization region (LIL) separate
from the high ionization region (HIL) where the Ly$\alpha$, CIV, CIII]
lines are formed (Collin-Souffrin et al. 1988). If the MgII line is
formed in a PIZ it is possible that the stronger BBB in NLSy1s will
push the ionization front further back into the cloud, resulting in a
smaller PIZ and weaker MgII emission, than for objects with a
``normal'' BBB.

However, the weaker MgII emission is not consistent with the stronger
FeII optical emission observed in NLSy1, which in photoionization
models is predicted to be formed in the same region (PIZ, Krolik \&
Kallman 1988, LIL, Collin-Souffrin et al. 1988).  This inconsistency
suggests that the FeII emission is instead generated in a different
region from MgII.  The observations of line variability in NGC~5548
(Sergeev et al. 1997) showed that the FeII optical multipets have a
very long time lag of several hundred days, while the MgII has a 30-50
day time lag, also implying that these lines are formed in different
regions. The FeII lines may be produced in the outer regions of the
accretion disk as suggested by Dumont \& Collin-Souffrin (1990) or in
a separate, mechanically heated region closely related to the compact
radio source as in Collin-Souffrin, Hameury \& Jolly (1988) (hence the
observed anti-correlation of FeII emission and radio flux).


\subsection{High luminosity to the Eddington luminosity ratio}

It has been suggested by a number of authors that NLSy1 galaxies as a
class have systematically higher ratios of their luminosity to the
Eddington luminosity, i.e. they have systematically lower masses in a
given luminosity range than Sy1 galaxies and QSOs (Pounds, Done \&
Osborne 1995, Wandel 1997). This suggestion was made based on the
analogy with the Galactic black hole candidates. We will address this
suggestion now.

\subsubsection{Continuum properties}

One of the current explanations of the soft-X-ray excess in AGN is
reprocessing of the hard X-rays by partially ionized, optically thick
matter, probably in the accretion disk. The model describes well the
soft X-ray continuum of low-luminosity, flat $\alpha_{softX}$ Seyfert
galaxies, but has problems with fitting the steepest $\alpha_{softX}$
spectra (see Fiore, Matt \& Nicastro 1997), which characterize NLSy1.
The steep $\alpha_{softX}$ can instead be explained by emission from the
innermost part of an accretion disk which is then
Comptonized by an optically thin, hot corona surrounding the disk
(Czerny \& Elvis 1987; Laor et al. 1997).

Theoretical models which can explain both the presence of the BBB and
the hard X-ray emission are based either on radial or horizontal
stratification between the hot optically thin and cold, optically
thick accretion flow (Wandel \& Urry 1991, Shapiro, Lightman \&
Eardley 1976, for a review see Wandel \& Liang 1991). In this paper we
use the model of an accretion disk corona (ADC) by Witt, Czerny \&
\. Zycki (1997), where the corona itself accretes and generates energy
through viscosity, and the division of the flow into optically thin
and optically thick regions results from the cooling instability
discussed by Krolik, McKee and Tarter (1981). Such a model is able to
predict the fraction of the energy generated in the corona instead of
adopting this quantity as a free parameter.  The model is fully
defined by 3 parameters: the mass of the central black hole
($M_{bh}$), the accretion rate or the ratio of the luminosity to the
Eddington luminosity ($L/L_{Edd}$) and the viscosity parameter
($\alpha_{vis}$, assumed to be the same in both the disk and the
corona).  The model predicts a systematic change in the opt/UV/X-ray
spectral energy distribution due to a change in $L/L_{Edd}$. A larger
ratio results in a more pronounced BBB, which is shifted towards
higher energies (resulting in stronger soft-X-ray emission and hence
steeper soft X-ray slopes).


We have determined continuum properties predicted by this model over a
large range of $L/L_{Edd}$ (0.001 to 0.7), $\alpha_{vis}$ (0.02 to
0.4) and black hole masses ($10^6$ to $10^{10} M_{\odot}$). We then
compared the observed continua of our NLSy1 with the UV luminosity at
2500\AA, and the soft and hard X-ray slopes ($\alpha_{softX}$ and
$\alpha_{hardX}$) predicted by the model. Table~5 shows the observed
$\alpha_{softX}$ (from ROSAT) and $\alpha_{hardX}$ (from ASCA) slopes
for each object, while Table~6 gives the best fitted model parameters
for each object.  Our model was able to reproduce the steep soft and
hard X-ray slopes within the observed uncertainties for most of the
NLSy1. However for two objects (IZw1, PKS~0558$-$504) we did not
succeed in fitting both the soft and hard-X-ray slopes
simultaneously. This may be due to the way we treat Comptonization in
our model (see Janiuk \& Czerny 1999 for further details).  In Fig.~5a,b
we show how the X-ray slopes change with the model parameters. Each
curve represents one value of $L/L_{Edd}$ and $\alpha_{vis}$ and a
full range of black hole masses, where smaller $M_{bh}$ lie at
smaller 2500\AA\ luminosities. We see clearly that only the large
ratios of $L/L_{Edd}$ can give the steep, observed soft X-ray slopes.

\begin{figure} [t!]
\vspace{9.0truecm}
{\includegraphics{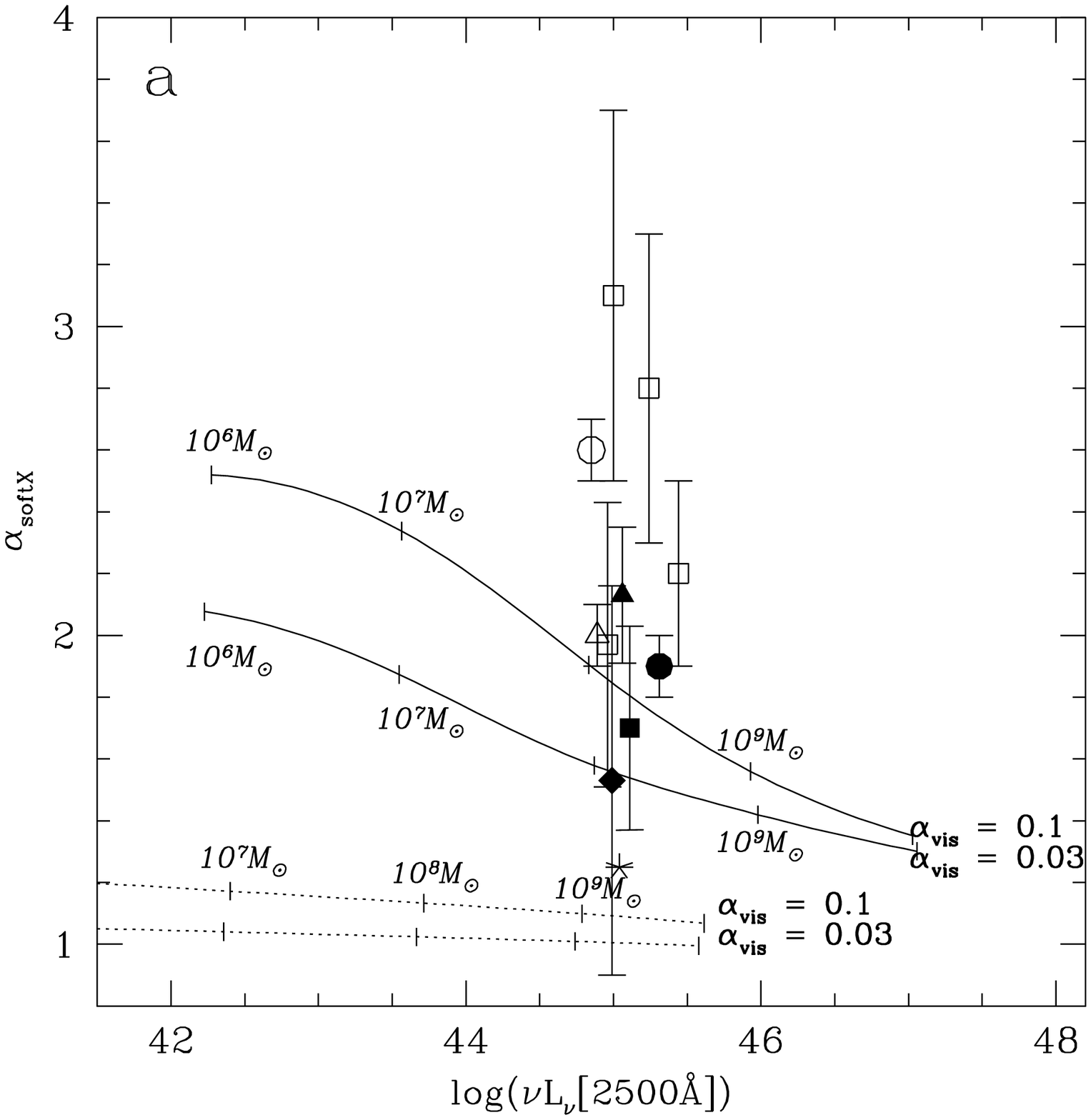}}
{\includegraphics{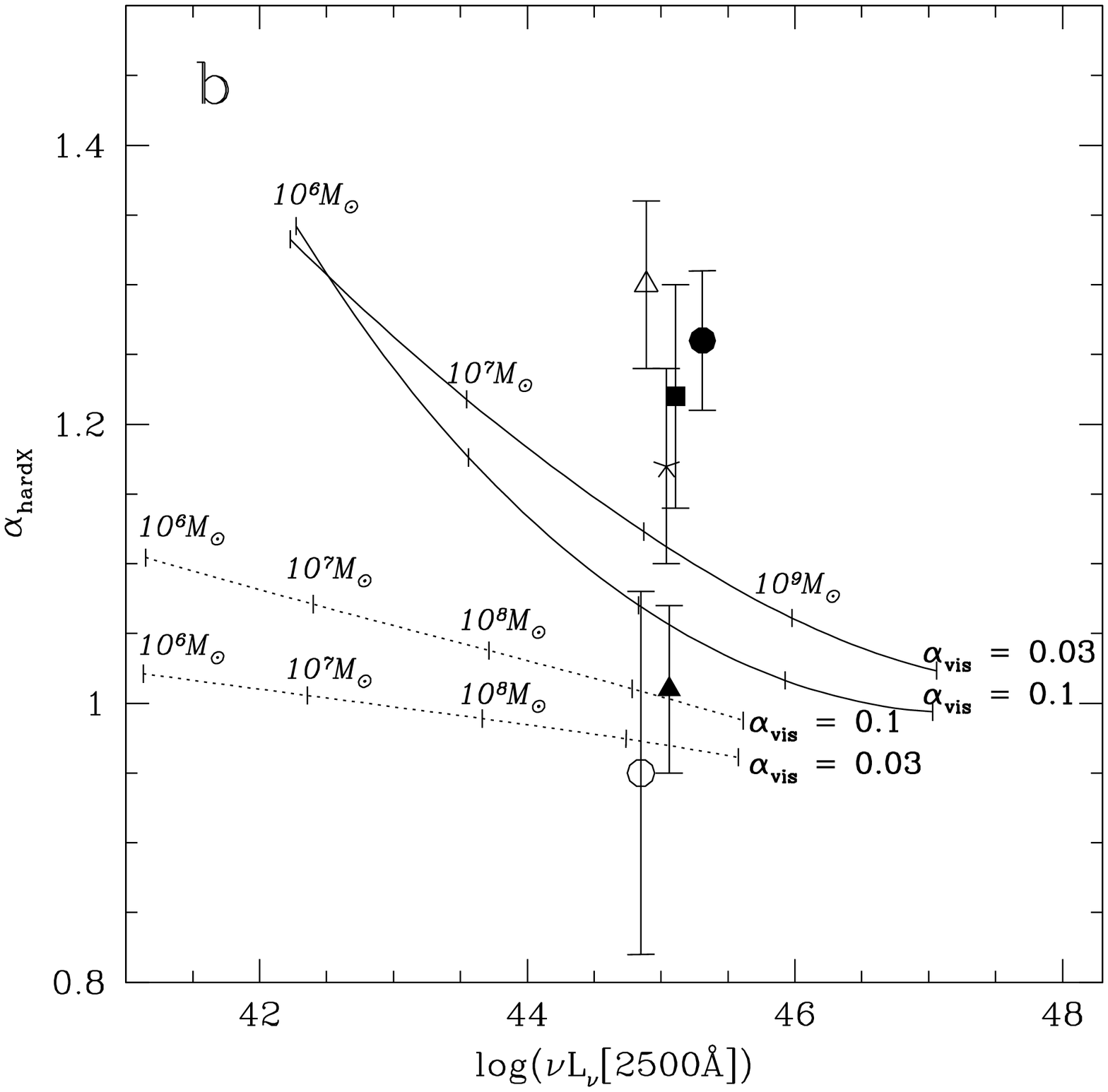}}
\figcaption{Comparison of the a) soft and b) hard X-ray
spectral slopes of NLSy1 objects in our sample with those 
predicted by the ADC model. Each curve represents one value of
$L/L_{Edd}$ and $\alpha_{vis}$ with varying black hole masses 
as labeled (smaller
$M_{bh}$ lie at the smaller 2500\AA\ luminosities). Dashed lines
denote small $L/L_{Edd}$=0.01, solid line  $L/L_{Edd}$=0.3. The value
of viscosity parameter are as labeled. 
Clearly the larger $L/L_{Edd}$ are needed to reproduce the steep soft
X-ray slopes of NLSy1s. The following symbols denote:
open triangle - IZw1, 
filled circle - PKS~0558$-$504, 
filled triangle - PG~1211+143, 
filled square - IRAS~13349+248, 
filled diamond - PG~1351+640, 
open circle - Mrk~478, 
star - IIZw136.}
\end{figure}


As has been shown by Czerny, Witt \& \.Zycki (1997), quasars radiate
usually at $\sim$ 0.01-0.2 of their Eddington luminosity, while
Seyfert galaxies radiate at $\sim$ 0.001-0.3. Our NLSy1 (where we use
the same ADC model as Czerny, Witt \& \.Zycki 1997 to fit the 
parameters of the central engine) radiate at
$L/L_{Edd}$ $\sim$ 0.27-0.58, much larger than the typical AGN. The
masses of the central black hole calculated from the model ($\sim 10^8
M_{\odot}$ to $10^{9} M_{\odot}$) for our objects are of the same
order as masses found in typical Seyfert~1 galaxies, but the
bolometric luminosities are larger, and comparable to those of QSOs
(see Wilkes et al. 1999, Table 12 for comparison).  This is deduced
from the stronger, higher energy BBB and
places the NLSy1 in a transition zone between the Sy1s and QSOs
i.e. among Sy1s with larger luminosities or QSOs with lower masses. We
note that, while the absolute numbers we deduce depend upon the
particular ADC model used, the general trends do not.

\subsubsection{Density and the radius of the BLR}

The structure and the dynamics of the BLR is complex, as suggested by
variability studies in the case of Seyfert galaxies. However, we can
analize the scaling properties of the whole BLR of an object with the
properties of the central source, including the shape of the X-ray
continuum.

The BLR gas Compton heated by the ionizing continuum will form (in any
geometry) two phases: a cool phase with $T_{c} \sim 10^4$ (the BLR
clouds) and a hot phase with $T_{h} \sim 10^8$ (the intercloud medium,
see Krolik and Kallman 1988, Czerny \& Dumont 1998, Wandel \& Liang
1991), when in equilibrium.  The precise values of these
temperatures depend on the shape of the continuum.

In the context of the two-phase model, 
we will now investigate how the properties of the BLR change
due to the steeper X-ray continuum of a NLSy1. We use the ionization 
parameter of Krolik, McKee \& Tarter (1981):
\begin{equation}
\Xi = \frac{2.3F_{ion}}{cp} = \frac{2.3 F_{ion}}{\frac{ck \rho_{c} T_{c}}{\mu H}}
\end{equation}
where $p$ is the total pressure,  $\rho_{c}$ and $T_{c}$  the density and
temperature of the cold phase, and 
$F_{ion}$ is the flux above 1 Ryd determined by the ionizing
luminosity of the central source $ L_{ion}$ and the current radius $r$
(where effects of geometry have been neglected):
\begin{equation}
F_{ion} = L_{ion} / 4 \pi r^{2}.
\end{equation}

The two phases coexist at a value of the ionization parameter,
$\Xi_h$, which scales with the hot phase temperature, $T_{h}$ in the
following way (Begelman et al. 1983):
\begin{equation}
\Xi_{h} = 0.65 \left( \frac{T_{h}}{10^8}\right)^{-3/2} 
\end{equation}

The BLR is most probably radially extended. For the purpose of
exploring the various dependencies, we determine a representative
radius for the BLR. Note that this is a scaling factor rather than 
the specific radius at which a particular emission line is generated. 
If the cloud number density profile is flatter than $r^{-2}$,
then most of the emission would come from the outer radii of the
BLR. As in the case of the Inverse Compton heated coronae discussed by
Begelman et al.  (1983), a nearly hydrostatic corona will exist up to
a radius where the temperature of the hot medium is equal to the
``escape'' temperature (i.e. the virial temperature). At larger radii
the corona is heated to temperatures exceeding the escape temperature,
becomes unstable and  forms  an outflowing wind. 
We therefore identify the outer edge of the BLR, $r_{BLR}$ with the radius
where the hot medium temperature is equal to the virial temperature
\begin{equation}
kT_{h} = \frac{GM_{bh}m_{H}}{r_{BLR}}
\end{equation}

The size of the BLR  expressed in units of the Schwarzschild 
radius, $R_{Schw}$ is then given by: 
\begin{equation}
r_{BLR}/R_{Schw} = {m_H c^2 \over 2 k T_h},
\end{equation}
so a lower value of the hot medium temperature in NLSy1 galaxies is
consistent with larger values of $r_{BLR}/R_{Schw}$ and, consequently,
lower values of the typical velocities.

A similar conclusion, that the BLR radius is larger in NLSy1s, was
reached by Wandel (1997) who assumed that the representative radius of
the BLR is determined by the requirement to have a standard value of
the ionization parameter. He then showed that the size of the BLR
region is dependent not only on the luminosity of the central source,
but also on the soft X-ray spectral slope. A steeper (softer) X-ray
spectrum has a stronger ionizing power and hence, for a constant
ionizing parameter, the BLR clouds are at larger distances from the
central source, have smaller velocity dispersions and as a result form
narrower emission lines. Laor et al. (1995) also reach a similar
conclusion but in their picture the narrow lines in NLSy1 result
purely from the lower black hole mass. In our scenario the lower black
hole mass and the shape of the SED (i.e. the steeper soft-X-rays which
decrease $T_{h}$) combine to produce the narrow lines.

Combining equations (1)-(4) we  estimate the cloud density:
\begin{equation}
\rho_{c} \sim \frac{L}{M_{bh}^{2}} \times \frac{T_{h}^{7/2}}{T_{c}}
\end{equation}

or using logarithms:

\begin{equation}
\log \rho_{c} \sim \log L - 2 \log M_{bh} + 7/2 \log T_{h} - \log T_{c}
\end{equation}

\noindent As has been argued in Section 3.2.1, NLSy1s have bolometric
luminosities comparable to QSOs, although their central black holes
have lower masses. The median value of a black hole in quasars is
$\sim$ $10^{10}M_{\odot}$ (see Czerny, Witt \& \.Zycki 1997
calculations), while the median black hole mass in NLSy1s (as inferred
from our calculations, using the same ADC model - see Table~6) is
$10^{8.26} M_{\odot}$ i.e.  $\sim$ 55 times lower.  Let us assume that
a typical quasar SED is composed of a power law and an accretion disk
spectrum peaking at 10 eV ($\log \nu = 15.38$, 1240\AA), while a
typical NLSy1 SED has a power law and a disk peaking at 80 eV (however
note that the most extreme NLSy1 RE~J1034$+$396 has its peak at 120eV
- see Puchnarewicz et al. 1995).  Krolik and Kallman (1988) calculated
the Compton temperatures of the hot phase for these SEDs, normalizing
both to have the same total ionizing energy. The 10eV bump spectrum
gave Compton temperatures $\sim$ 3.0 $\times 10^{7} K$ while the 80eV
bump gave a lower temperature $\sim$ 8.0 $\times 10^{6} K$. At the
same time the temperature of the cool phase increased by a factor
$\sim$ 3.0 (0.5 in logarithm see Krolik \& Kallman 1988 Fig. 2). Hence
the Compton temperature of the hot phase in NLSy1 and QSOs differs by:
$\log T_{h, NLSy1} - \log T_{h, QSO} = -0.57$ and the temperature of
the cold phase is larger by: $\log T_{c, NLSy1} - \log T_{c, QSO} =
0.5$.  Substituting the above values into equation (7) implies that
$\log \rho_{c, NLSy1} - \log \rho_{c,QSO} \approx 1$ i.e the densities
of the BLR should be higher by a factor of 10 in NLSy1 than in typical
QSOs with redder BBB.


The larger BLR radii and larger by a factor 10 densities obtained 
from our modeling (as being due to hotter BBBs)
are consistent with the narrow lines and line ratios observed 
in NLSy1s. Thus we conclude that the unusually hot and strong BBB in
NLSy1s can naturally produce their observed UV spectra.



\section{NLSy1 vs. BAL QSOs}

It has been suggested (e.g. Leighly et al. 1997, Lawrence et al. 1997) that
there may exist a connection between NLSy1 and BAL QSOs. Both these
classes have strong FeII\,$\lambda$4570 and AlIII\,$\lambda$1857 emission
and weak CIV\,$\lambda$1549 and [OIII$]\lambda$5007. Their continua are
red in the optical and strong in the IR; additionally both classes are
mostly radio-quiet.  Leighly et al. (1997) reported evidence for
relativistic outflows in three NLSy1.

Observationally there are also many differences. NLSy1 are strong soft-X-ray
emitters, while BAL QSOs are weak, possibly due to X-ray absorption
(Mathur, Elvis \& Singh 1995). BAL QSOs are thought to be seen more
edge-on, at viewing angles skimming the edge of the dusty torus
(Turnshek et al. 1996, Aldcroft, Elvis \& Bechtold 1993). NLSy1s,
on the other hand, are probably viewed more face-on, as they show low
absorption from the torus (Boller, Brandt \& Fink 1996) and some even
show beaming in their radio spectra (e.g. PKS~0558-504, Remillard et
al. 1991).

In high resolution HST spectra NLSy1 show absorption features
which are much weaker than in BAL QSOs (see Table~2). However this is
expected since 50\% of Seyfert 1s show absorption features (Crenshaw
et al. 1995).  The optical spectra of some BAL QSOs may resemble the
spectra of NLSy1, showing narrow H$\beta$ with FWHM $<$
2000~kms$^{-1}$ (that is why the two BAL QSOs: PG~1351+640 and
PG~1411+442 were initially chosen to be in our sample), but this only
cautions us that basing classifications on optical spectra alone is
potentially misleading.

%
%
%
%

\section{Conclusions}

In this paper we have studied the UV emission line properties of a
class of extreme opt/UV/X-ray AGN: the narrow line Seyfert 1
galaxies. We found 11 NLSy1s that had been observed in the UV by
either HST or IUE. We have shown that in comparison with ``normal''
broader line AGN, the equivalent widths of CIV and MgII are
significantly smaller (NLSy1 have EW(CIV)$<$60 and 
EW(MgII)$<$20, normal AGN have EW(CIV)$<$210 and
EW(MgII)$<$120), the EW of AlIII larger (few \AA), and the UV line widths
are narrower (although not as narrow as the optical H$\beta$
line). Also the CIII]/Ly$\alpha$, CIV/Ly$\alpha$ and MgII/Ly$\alpha$
line ratios are smaller, while those of SiIII]/CIII], SiIV+OIV]/CIV
lines are larger. Photoionization models predict that these line
ratios are formed in material with densities higher, by a factor few
(less than 10) than standard BLR cloud densities, and with the
ionization parameter lower by a factor 10. These parameters however
predict higher MgII/Ly$\alpha$ ratio, in contradiction to the lower
ratios observed requiring that MgII be produced in a separate region.

We have fitted the SEDs of our NLSy1s to the Witt, Czerny \& \.Zycki
(1997) model of an accretion disk with a Compton cooled corona and
found that NLSy1s radiate at $0.27 < L/L_{Edd} < 0.58$, much larger
than the typical AGN ($L/L_{Edd} <$ 0.3). The masses of the central
black holes calculated from the model are, in our objects, of the
order of masses found in typical Seyfert 1 galaxies ($10^8 M_{\odot}$)
but the bolometric luminosities ($\nu L_{\nu} \sim 10^{46}$ erg
s$^{-1}$) are larger and comparable to those of QSOs.

Krolik \& Kallman (1988) predict that steeper soft-X-ray BBBs, such as
these of NLSy1s, change the equilibrium of the two-phase
cloud-intercloud medium, decreasing the temperature of the hot
intercloud medium (which we assume to be the corona above the
accretion disk) and increasing the temperature of the cool BLR
clouds. We show that this change in equilibrium increases the density
of the BLR clouds resulting in a change of the observed line
intensities and ratios consistent with these in NLSy1s. In addition
the resulting decrease in $T_h$, causes an increase in the radius of
the BLR, a correspondingly lower velocity dispersion and narrower
lines as observed in NLSy1s.

The NLSy1s lie at the extreme end of the Boroson and Green eigenvector
1 (Boroson \& Green 1992), which was then found (Brandt \& Boller
1998) to link the soft X-ray properties with the optical properties
i.e. the FeII/H$\beta$ and $[$OIII] strengths and H$\beta$ line
width. We have found that the NLSy1s have very weak CIV and CIII]
lines, and narrow UV lines extending the set of parameters linked to
eigenvector 1. The large BLR cloud densities, deduced from these
characteristic UV line ratios, are probably due to the steep soft
X-ray SEDs, which are in turn, the result of larger $L/L_{Edd}$ ratios
(as inferred from the Witt, Czerny \& \.Zycki 1997 ADC model). In this
scenario a larger $L/L_{Edd}$ is the physical parameter driving the
Boroson \& Green eigenvector 1.


\acknowledgments

We are grateful to Niel Brandt for helping us to obtain a complete
list of known NLSy1 and their X-ray slopes, Adam Dobrzycki for
providing us with the HST data and Ken Lanzetta for the IUE Atlas of
AGN spectra. We wish to thank Martin Gaskell, Martin Elvis, Marianne
Vestergaard, Kirk Korista and Suzy Collin-Souffrin for valuable
discussions and thank the anonymous referee for comments that improved
the manuscript. JK greatfully acknowledges the support of a
Smithsonian pre-doctoral fellowship at the Harvard-Smithsonian Center
for Astrophysics and grant no. 2P03D018.16 of the Polish State
Committee for Scientific Research (JK and BCz). BJW acknowledges NASA
contract NAS8-39073 (Chandra X-ray Center) and SM a NASA grant
NAG5-3249 (LTSA)

\begin{deluxetable}{llllc}
\tablenum{1}
\tablecaption{Sample}
\tablehead{
\colhead{Name} & 
\colhead{$\alpha$(J2000)} &
\colhead{$\delta$(J2000)} & 
\colhead{z} & 
\colhead{$\log \nu L_{\nu}$} \\
\colhead{} &
\colhead{} &
\colhead{} &
\colhead{} &
\colhead{2500\AA$^a$} 
}
\startdata

IZw1* &00 53 34.94 & $+$12 41 36.2 & 0.0611 &44.89 \\

     E 0132$-$411&01 34 57.36&$-$40 56 22.4& 0.266 & 45.00\\

NAB 0205+024  & 02 07 49.86&$+$02 42 55.9& 0.1564 &45.24\\

PKS 0558$-$504&05 59 47.37 &$-$50 26 51.8  & 0.137 & 45.31 \\

PG 1211+143*&12 14 17.60 &+14 03 12.5   & 0.085 &45.06\\

IRAS~13349+2438&13 37 18.73 & +24 23 03.3  & 0.107 & 45.11\\
PG 1351+640$^{BAL}$ & 13 53 15.78 & +63 45 44.8 & 0.087 & 44.99 \\


PG 1411+442*$^{BAL}$ &14 13 48.39&+44 00 13.6  & 0.090 & 44.96 \\

MRK 478*&14 42 07.46 & +35 26 22.9 & 0.0781& 44.85 \\

PG 1444+407* &14 46 45.95 &+40 35 06.0 & 0.267 & 45.44 \\



IIZw136 & 21 32 27.81 & +10 08 19.5 & 0.061 & 45.04  \\
\enddata

\tablenotetext{}{* HST spectrum}
\tablenotetext{}{{\it BAL}: also classified as a broad-absorption line
BAL QSO}
\tablenotetext{}{a: the luminosity at
2500\AA\ has been obtained by extrapolating B (or V) magnitude and assuming 
a continuum slope of 0.5 ($H_{0} = 50$ km s$^{-1}$ Mpc$^{-1}$)}

\end{deluxetable}

\clearpage

\begin{center}
\leavevmode
\epsfxsize=6.1in
\epsfysize=3.7in
\epsfbox{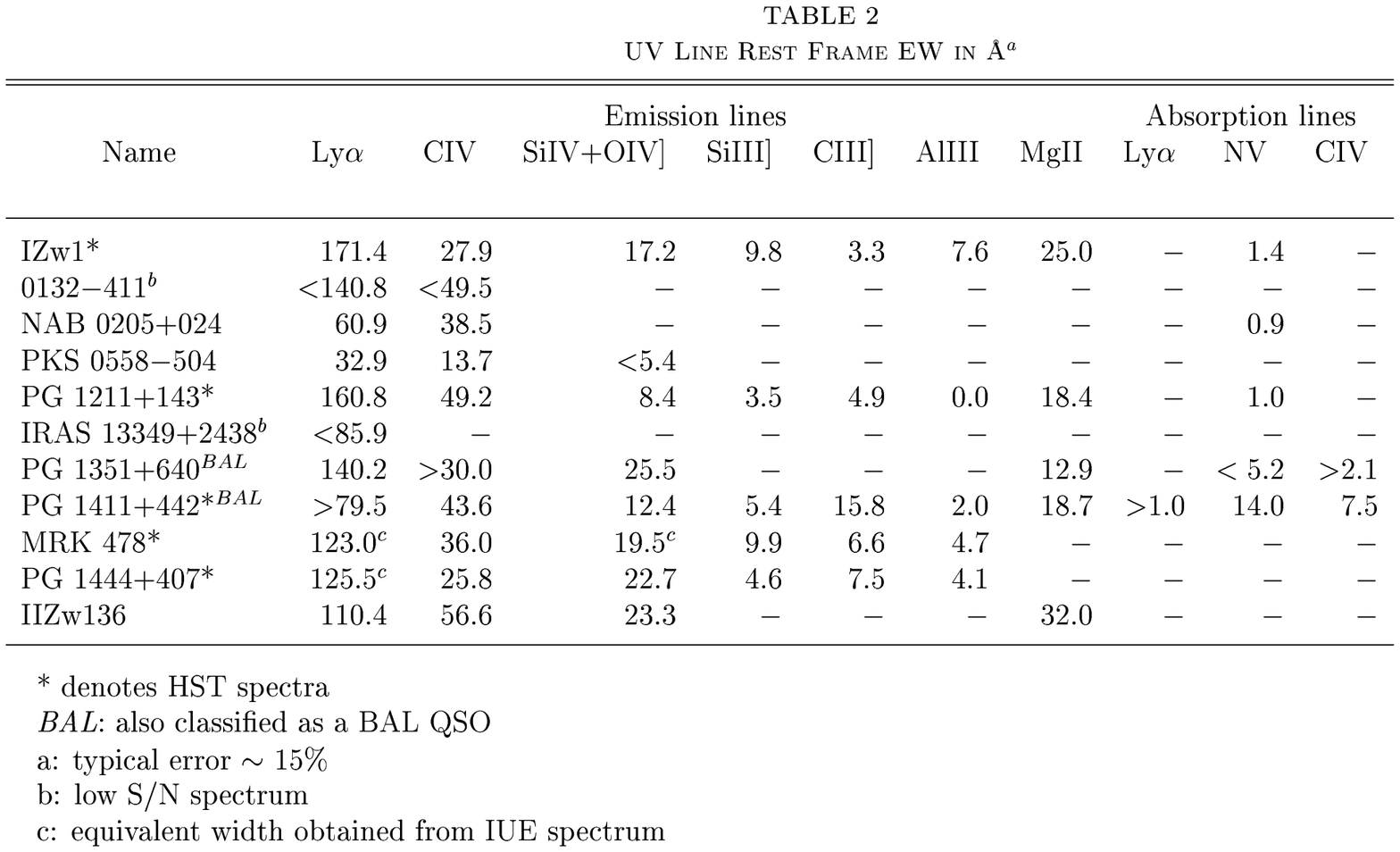}
\end{center}
\eject

\begin{center}
\leavevmode
\epsfxsize=6.1in
\epsfysize=4.9in
\epsfbox{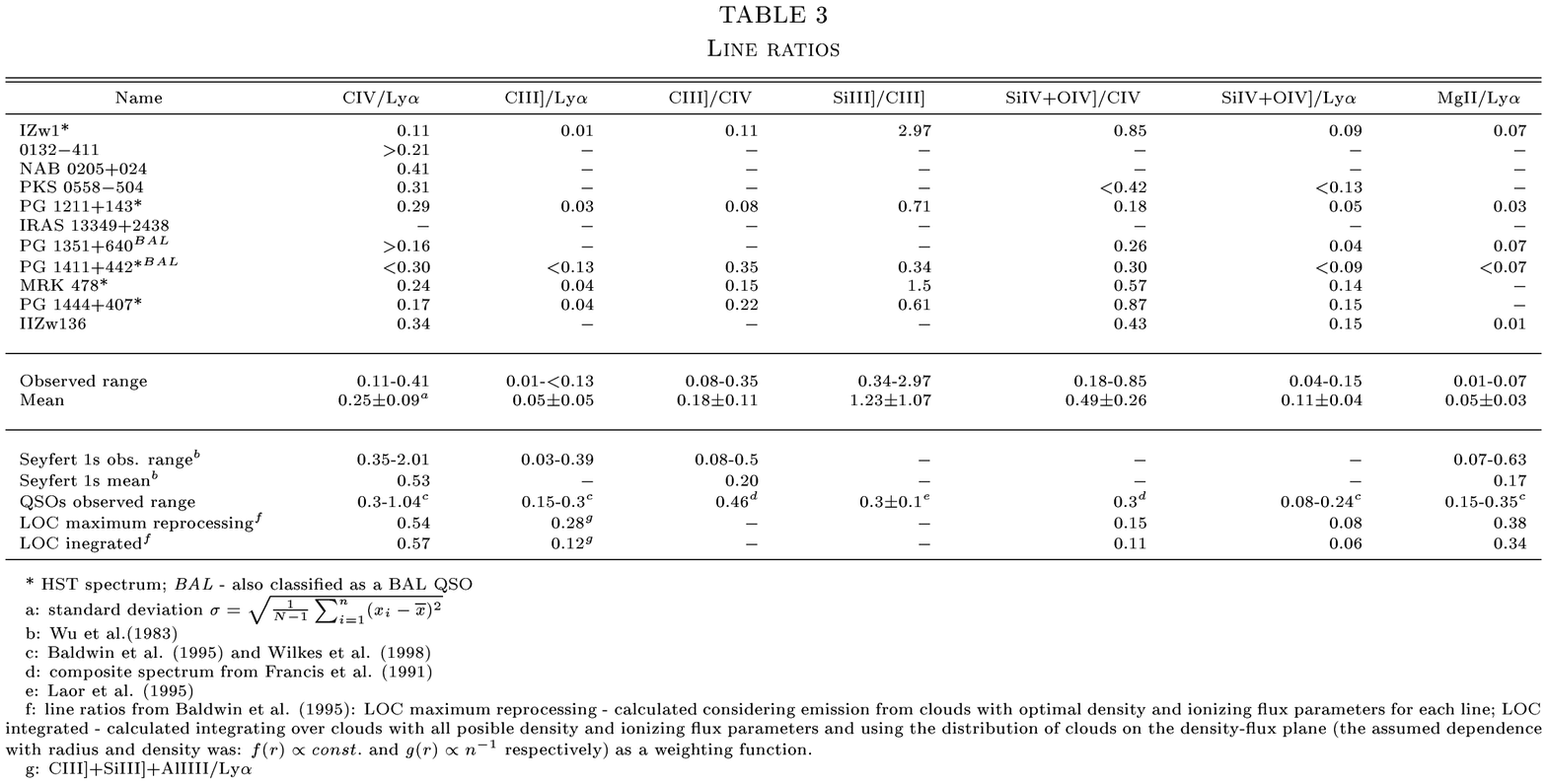}
\end{center}
\eject

\begin{deluxetable}{lcccc}
\tablenum{4}
\tablecaption{Line widths in km/s}
\tablewidth{0pt}
\tablehead{
\colhead{Name} & 
\colhead{Ly$\alpha$} & 
\colhead{CIV} &
\colhead{MgII} &
\colhead{H$\beta^a$}
}
\startdata
IZw1*               & 1730  &3190  & 1850 & 1240\\
0132$-$411$^a$         & $<$3920 & $<$2460   & - &1930\\
NAB 0205+024       & 1630 & 2520  & -&1100 \\
PKS 0558$-$504     & 3820 & 3540  & -&1500 \\
PG 1211+143*        & 1600 & 1940  & 1945&1900 \\ 
IRAS 13349+2438$^b$  & $<$7000 &-  & -&2100\\
PG 1351+640$^{BAL}$        & 2370 & $>$2110 & 2140&860 \\
PG 1411+442*$^{BAL}$       & $>$1850 & 2300  & 1780&2670 \\
MRK 478*            & 2810 & 2820  & -&1370 \\
PG 1444+407*        & 3700 & 6180  &-&2480\\
IIZW136            & 2130 & 2363 & 1980&2060 \\ 
		   & & & & \\ \tableline
&&&& \\
Mean		   & 2960$\pm$1611 & 2942$\pm$1237 & 1939$\pm$133
		   &1746$\pm$577 \\
		   & & & & \\ \tableline
&&&&\\
QSOs$^c$ & 5399$\pm$2757 & 4793$\pm$1765 & 5566$\pm$2426& \\
QSOs$^d$  &$-$ &5150$\pm$1680&4580$\pm$1890& \\
QSOs$^e$ & 3454$\pm$1291 & 4335$\pm$1550 & 3774$\pm$2212& \\
\enddata
\tablenotetext{}{* HST spectra; {\it BAL} - also classified as a BAL QSO}
\tablenotetext{}{a: after Boller, Brandt, Fink (1996)}
\tablenotetext{}{b: very low S/N spectra}
\tablenotetext{}{c: Wilkes et al. (1998)}
\tablenotetext{}{d: Baldwin, Wampler \& Gaskell (1986)}
\tablenotetext{}{e: Corbin \& Boroson (1996)}
\end{deluxetable}

\clearpage

\begin{deluxetable}{llllr}
\tablenum{5}
\tablecaption{Observed soft and hard X-ray indices}
\tablehead{
\colhead{Name} & 
\colhead{$\alpha (0.1-2.5keV)$} &
\colhead{Ref.} & 
\colhead{$\alpha (2-10keV)$} &
\colhead{Ref.}
}
\startdata
IZw1               &     2.0$\pm$0.1   &1 & 1.3$\pm$0.06&7 \\
0132$-$411         &     3.1$\pm$0.6   &1 & $-$&$-$\\
NAB 0205+024       &     2.8$\pm$0.5   &2 & 1.09$\pm$0.10&10 \\
PKS 0558$-$504     &     1.9$\pm$0.1       &3 & 1.26$\pm$0.05&3 \\
PG 1211+143        &     2.13$\pm$0.22 &4 & 1.01$\pm$0.06&8 \\
IRAS 13349+2438    & 	 1.70$\pm$0.33     &5 & 1.22$^{+0.08}_{-0.08}$&8 \\
PG 1351+640$^{BAL}$&     1.53$\pm$0.63     &4 & $-$&$-$\\
PG 1411+442$^{BAL}$&     1.97$\pm$0.46 &4 & $-$ &$-$ \\ 
MRK 478            &     2.6$\pm$0.1   &1 & 0.95$\pm$0.13&8 \\
PG 1444+407        &     2.2$\pm$0.3   &1 & $-$&$-$ \\
IIZw136            &     1.25$\pm$...      &6 & 1.17$\pm$0.07$^a$&9\\
\enddata


\tablenotetext{}{{\it BAL} - also classified as a BAL QSO}
\tablenotetext{}{a: slope from {\it Ginga} determined between 2-18keV
(ref. 9)}
\tablenotetext{}{{\bf References:} \\
1 - Boller, Brandt \& Fink (1996), \\ 
2 - Fiore et al. (1995), \\
3 - Brandt, private communication, \\ 
4 - Wang, Brinkmann \& Bergeron (1996), \\
5 - Brandt, Fabian \& Pounds (1996), \\
6 - Wang, Lu \& Zhou (1998) (error on $\alpha (0.1-2.5keV)$ not
available), \\
7 - Hayashida (1997), \\ 
8 - Brandt, Mathur \& Elvis (1997), \\ 
9 - Lawson \& Turner (1997),  \\
10 - Fiore et al. (1998) \\
}
\end{deluxetable}

\clearpage

\begin{deluxetable}{lclccc}
\tablenum{6}
\tablecaption{Soft and hard X-ray indices from best fit model}
\tablehead{
\colhead{Name} & 
\colhead{$L/L_{Edd}$}&
\colhead{$\alpha_{vis}$} &
\colhead{$\log M_{bh}$}&
\colhead{$\alpha_{softX} ^{c}$} & 
\colhead{$\alpha_{hardX} ^{d}$}
}
\startdata
IZw1$^b$               &0.58&0.027&7.93&2.02&1.17\\
0132$-$411$^a$         &0.27&0.30&8.29&2.52&1.05\\ 
NAB 0205+024       &0.27&0.30&8.45&2.50&1.04\\
PKS 0558$-$504$^b$     &0.58&0.027&8.89&1.89&1.14\\
PG 1211+143        &0.30&0.14&8.26&2.02&1.05\\
IRAS 13349+2438    &0.58&0.03&8.12&1.99&1.15\\
PG 1351+640$^{BAL,a}$        &0.27&0.03&8.20&1.51&1.11\\
PG 1411+442$^{BAL, a}$        &0.58 &0.027&7.96&2.00&1.17\\
MRK 478            & 0.27&0.3 &8.12 &2.53&1.07\\
PG 1444+407$^a$        &0.27&0.3&8.61&2.49&1.04\\
IIZw136            &0.27&0.03&8.20&1.51&1.11\\
\enddata
\tablenotetext{}{{\it BAL} - also classified as a BAL QSO}
\tablenotetext{}{a: these objects do not have observed hard-X-ray
slopes, hence the derived model parameters are not well constrained}
\tablenotetext{}{b: we could not fit both the soft and  hard X-ray
slopes simultaneously}
\tablenotetext{}{c: soft X-ray index measured from 0.1-2.5 keV
corresponding to a ROSAT slope}
\tablenotetext{}{d: hard X-ray index measured from 2-10 keV
corresponding an ASCA slope} 
\end{deluxetable}

\end{document}